\documentclass[11pt,twoside]{article}

\usepackage{asp2006}
\usepackage{epsf}
\usepackage{lscape}
\usepackage{graphicx}

\begin{document}
\title{Evidence for Nonlinear Diffusive Shock Acceleration of Cosmic-Rays in the 2006 Outburst of RS Ophiuchi}   
\author{V. Tatischeff\altaffilmark{1} and M. Hernanz}
\affil{Institut de Ci\`encies de l'Espai (CSIC-IEEC), Campus UAB,\\ Fac. Ci\`encies, 08193 Bellaterra, Barcelona, Spain}

\altaffiltext{1}{Permanent adress: CSNSM, IN2P3-CNRS and Univ 
Paris-Sud, F-91405 Orsay Cedex, France}

\begin{abstract} 
Spectroscopic observations of the 2006 outburst of RS Oph at both infrared (IR) and X-ray wavelengths have shown that the blast wave has decelerated at a higher rate than predicted by the standard test-particle adiabatic shock-wave model. The observed blast-wave evolution can be explained, however, by the diffusive shock acceleration of particles at the forward shock and the subsequent escape of the highest energy ions from the acceleration region. Nonlinear particle acceleration can also account for the difference of shock velocities deduced from the IR and X-ray data. We discuss the evolution of the nova remnant in the light of efficient particle acceleration at the blast wave. 
\end{abstract}

\section{Introduction}

X-ray observations of the latest outburst of RS Oph \citep{sok06,bod06} have allowed to clearly identify the forward shock wave expanding into the red giant (RG) wind and to estimate the time evolution of its velocity, $v_s$, through the well-known relation for a test-particle strong shock:
\begin{equation}
v_s = \bigg({16 \over 3} {k T_s \over \mu m_H}\bigg)^{0.5}~,
\end{equation}
where $k$ is the Boltzmann constant, $T_s$ is the measured postshock temperature and $\mu m_H$ is the mean particle mass. The X-ray emission has revealed that after an ejecta-dominated, free expansion stage lasting $\sim$6~days, the remnant rapidly evolved to display behavior characteristic of a shock experiencing significant radiative cooling. The lack or the very short duration of an adiabatic, Sedov-Taylor phase differs from the remnant evolution model developed after the 1985 outburst \citep[see][and references therein]{obr92}. 

The time-dependence of shock velocity has also been measured by IR spectroscopy \citep{das06,eva07}. Although the general behavior of the shock evolution was found to be consistent with that deduced from the X-ray emission, the shock velocities determined from the IR data are significantly greater than those obtained using equation~(1) together with the X-ray measurements of $T_s$ (see  Fig.~1a). 

We have shown that production of nonthermal particles by diffusive shock acceleration at the blast wave can account for these observations \citep{tat07}. We have proposed that the difference of shock velocities deduced from the IR and X-ray data is due to the use of equation~(1), which is known to underestimate $v_s$ when particle acceleration is efficient \citep{dec00,ell07}. The observed early cooling of the blast wave is explained by the escape of high-energy particles from the acceleration region, which we found to be much more effective to cool the shock than radiative losses. Thus, nonlinear diffusive shock acceleration has important implications for the evolution of the nova remnant. 

\section{The density in the red giant wind}

The RG wind density can be estimated from both the photoelectric absorption of the postshock X-rays and the free-free absorption of the radio synchrotron emission, which presumably arises from electron acceleration at the blast wave. Using the absorbing column density $N_H$ measured with the {\it Swift} X-Ray Telescope \citep{bod06}, and assuming the abundances in the RG wind to be solar (see Wallerstein et al., these proceedings), we obtain from equation~(4) of \citet{bod06}: $\dot{M}_{\rm RG} / u_{\rm RG} \approx 4 \times 10^{13}$~g~cm$^{-1}$, where $\dot{M}_{\rm RG}$ and $u_{\rm RG}$ are the RG mass-loss rate and wind terminal speed, respectively. It is noteworthy that the values of $N_H$ are still uncertain, because they were determined by rough single-temperature 
fits \citep{bod06}.
 
To calculate the optical depth to radio free-free absorption, we use the formula derived by \citet{obr92}, which allows for the effects of H recombination in the stellar wind. For $\dot{M}_{\rm RG} / u_{\rm RG} = 4 \times 10^{13}$~g~cm$^{-1}$ and a uniform wind temperature $T_{W}$ in the range (1--3)$\times 10^4$~K, we find the time of free-free optical depth unity at 6.03~GHz to be 20--30~days. This result is consistent with the broad peak of radio emission observed with the Very Large Array (VLA) and MERLIN after $t=20$~days, but is inconsistent with the early rapid rise of radio emission detected at days 4 and 5 after outburst (Eyres et al., these proceedings). If this early emission is associated with the blast wave expanding at $v_s\sim 4300$~km~s$^{-1}$ (Fig.~1a), it implies $\dot{M}_{\rm RG} / u_{\rm RG} < 7 \times 10^{12}$~g~cm$^{-1}$. From GMRT observations at frequencies $<1.4$~GHz, \citet{kan07} obtained $\dot{M}_{\rm RG} / u_{\rm RG} = 5 \times 10^{12}$~g~cm$^{-1}$. However, the X-ray- and radio-based estimates of the RG wind density could be reconciled if the first early peak of radio emission is associated with a synchrotron jet expanding much faster than the blast wave (see O'Brien et al. and Rupen et al., these proceedings). 

\section{Properties of the Cosmic-Ray Modified Shock}

The effects of particle acceleration at the blast wave depend on the strength of the magnetic field just ahead of the shock, $B_0$. Assuming the magnetic field in the RG wind to be in equipartition with the thermal energy density, we have $B_0 = \alpha_B (2 \dot{M}_{\rm RG} kT_W / u_{\rm RG} r_s^2 \mu m_H)^{0.5}$, where $\alpha_B$ is a factor accounting for a possible amplification of the magnetic field in the shock region and $r_s$ is the shock radius. Here and in the following, we adopt $\alpha_B=1$, $\dot{M}_{\rm RG} / u_{\rm RG} = 2 \times 10^{13}$~g~cm$^{-1}$ and $T_{W}=3\times 10^4$~K, which give $B_0 = 0.06$~G at $t=t_1$. With such values of $B_0$, we have shown \citep{tat07} that accelerated electrons and protons can have achieved TeV energies in few days after outburst.
 
To study the modification of the shock structure induced by the backreaction of the energetic ions, we use the model of nonlinear diffusive shock acceleration developed by \citet{ber99}, with the prescription of \citet{bla05} for the particle injection. Calculated temperatures of the postshock gas are shown in Fig.~1b for two values of the parameter $\eta_{\rm inj}$, which is the fraction of postshock thermal protons injected into the acceleration process. The other parameters of the model are as in \citet{tat07}. We see that the temperatures measured with {\it RXTE} and {\it Swift} can be well reproduced with $\eta_{\rm inj}$=1.5$\times$10$^{-4}$. For $\eta_{\rm inj}$=10$^{-5}$, the test-particle approximation applies and the standard relation between $v_s$ and $T_s$ (eq.~[1]) overestimates the temperature. The model also provides the shock compression ratio and acceleration efficiency. We find that the energy carried off by high-energy particles escaping the shock region represents 10\%--20\% of the total energy flux at $t>t_1$.  

\begin{figure}[t]
\plotone{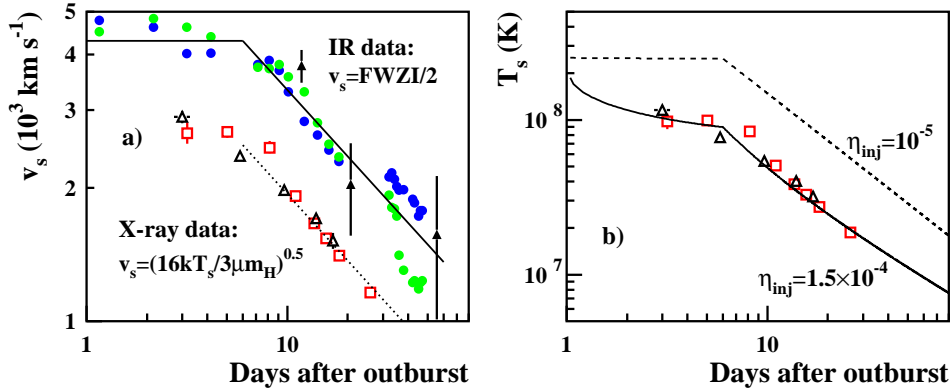}
\caption{(a) Time-dependence of forward shock velocity as deduced from FWZI of IR emission lines ({\it filled circles}: Das et al. 2006, {\it filled triangles}: Evans et al. 2007) and X-ray measurements of the postshock temperature ({\it open triangles}: Sokoloski et al. 2006, {\it open squares}: Bode et al. 2006). The IR data can be modeled by ({\it solid line}) $v_s(t) = 4300 (t/t_1)^{\alpha_v} {\rm~km~s^{-1}}$, where $t_1=6$~days and $\alpha_v=0$ (-0.5) for $t \leq t_1$ ($t>t_1$). (b) Calculated postshock temperature for two values of $\eta_{\rm inj}$ compared to the {\it RXTE} and {\it Swift} data.  
\label{fig1}}
\end{figure}
 
\section{Calculated X-Ray Fluxes}

\begin{figure}[t]
\begin{center}
\includegraphics[width=7.8cm]{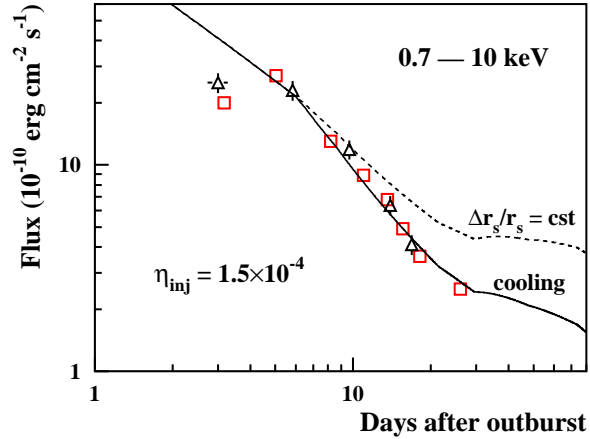}
\caption{Time-dependence of absorption-corrected X-ray flux in the 0.7--10 keV energy range. Using the MEKAL plasma emission model, corrections ranging from 0.76 to 0.90 were applied to the {\it RXTE} PCA fluxes, which are given by \citet{sok06} in the 0.5--20 keV energy range. {\it Dashed curve}: $\alpha_r=0$; {\it solid curve}: $\alpha_r=\alpha_v$ (see text). 
\label{fig2}}
\end{center}
\end{figure}

We compare in Fig.~2 calculated X-ray fluxes with the {\it RXTE} and {\it Swift} data of \citet{sok06} and \citet{bod06}, respectively. Using theoretical emissivities from the MEKAL plasma emission model and calculated values of $T_s$ and the postshock density from the nonlinear shock acceleration model, we determine the characteristic emission volume, $V(t) \cong 4 \pi r_s^2(t) \Delta r_s(t)$, required to reproduce the measured fluxes. We assume the relative thickness of the X-ray emitting shell to be of the form
$
{\Delta r_s(t) / r_s(t)}=f_r ({r_s(t) / r_s(t_1)})^{\alpha_r}~,
$
where $f_r$ and $\alpha_r$ are free parameters. The two curves shown in Fig.~2 are for $\alpha_r=0$ and $\alpha_r=\alpha_v$. The first value corresponds to an adiabatic expansion of the shock wave. It is clear that the observed flux evolution implies a shrinking of the relative shell thickness at $t>t_1$, as expected from a cooling shock. The normalization of the calculated curves to the data gives $f_r=6.8\% \times (D_{\rm kpc}/1.6)^2$, where $D_{\rm kpc}$ is the source distance in kpc. This result is consistent with hydrodynamic models of supernova remnant (SNR) evolution with efficient cosmic-ray acceleration \citep{ell07}, as well as with SNR observations \citep[e.g.][]{war05}. We note that a value of $\dot{M}_{\rm RG} / u_{\rm RG} < 7 \times 10^{12}$~g~cm$^{-1}$, as needed if the early radio emission is associated with the blast wave (see \S~2), would require a too large X-ray emission volume to reproduce the measured fluxes ($F_X \propto V (\dot{M}_{\rm RG} / u_{\rm RG})^2$). This gives some support for the assumption that the early radio peak is due to a fast synchrotron jet. 

\acknowledgements This work has been partially supported by the grants AGAUR 2006-PIV-10044 and MEC AYA2004-06290-C02-01.


\end{document}